\documentclass[aps,twocolumn,superscriptaddress]{revtex4-2}

\usepackage{amsmath,charter,multirow}
\usepackage{graphicx,longtable,xfrac}
\usepackage[colorlinks=true,citecolor=blue,linkcolor=blue,breaklinks=true]{hyperref}
\usepackage{amssymb}
\usepackage{amsmath}
\usepackage{color}
\usepackage{bm}
\usepackage{soul}
\usepackage{indentfirst}

\begin{document}

\title{Magnetic order in the \textit{S}\textsubscript{eff} = 1/2 triangular-lattice compound NdCd$_3$P$_3$}

\author{Juan R. Chamorro}
\altaffiliation{These authors contributed equally to this work.}
\affiliation{Materials Department, University of California, Santa Barbara, California 93106, USA}

\author{Azzedin R. Jackson}
\altaffiliation{These authors contributed equally to this work.}
\affiliation{Materials Department, University of California, Santa Barbara, California 93106, USA}

\author{Aurland K. Watkins}
\affiliation{Materials Department, University of California, Santa Barbara, California 93106, USA}

\author{Ram Seshadri}
\affiliation{Materials Department and Materials Research Laboratory, University of California, Santa Barbara, California 93106, USA}

\author{Stephen D. Wilson}
\email[]{stephendwilson@ucsb.edu}
\affiliation{Materials Department, University of California, Santa Barbara, California 93106, USA}

\date{\today}

\date{\today}

\begin{abstract}
We present and characterize a new member of the \textit{R}Cd$_3$P$_3$ (\textit{R}= rare earth) family of materials, NdCd$_3$P$_3$, which possesses Nd$^{3+}$ cations arranged on well-separated triangular lattice layers. Magnetic susceptibility and heat capacity measurements demonstrate a likely \textit{S}\textsubscript{eff} = 1/2 ground state, and also reveal the formation of long-range antiferromagnetic order at $T_{N} = 0.34$ K. Via measurements of magnetization, heat capacity, and electrical resistivity, we characterize the electronic properties of NdCd$_3$P$_3$ and compare results to density functional theory calculations. 

\end{abstract}

\maketitle

\section*{Introduction}

Crystalline compounds possessing geometric magnetic frustration remain at the forefront of research in condensed matter physics, owing to the plethora of interesting ground states they may harbor. Confinement in low-dimensional cases, such as quasi-one-dimensional systems and two-dimensional structures, can often result in enhanced quantum fluctuations capable of rendering unique quantum ground states, such as the quantum spin liquid state \cite{Balents2010,Broholm2020,Chamorro2021}. Materials systems possessing two-dimensional triangular lattices have, in particular, attracted much attention owing to the large degree of geometric frustration ensured by a two-dimensional array of edge-sharing triangles of magnetic moments, which further enhances quantum fluctuations and increases the likelihood of unconventional quantum ground states. Indeed, several triangular lattice quantum spin liquid candidates have been identified, such as those with transition metal cations such as NaRuO$_2$ \cite{Ortiz2023}, and rare-earth cations such as NaYbSe$_2$ \cite{Jia2020,Liu2018,Zhang2021}, YbMgGaO$_4$ \cite{Paddison2017,Li2019,Zhu2017}, and NaYbO$_2$ \cite{Bordelon2019, Ding2019,Bordelon2020}. 

The layered \textit{R}\textit{M}$_3$\textit{Pn}$_3$ (\textit{R}= rare earth; \textit{M} = Zn, Cd; \textit{Pn} = P, As) family of pnictide compounds is one that hosts \textit{R}$^{3+}$ cations on well-separated triangular lattice layers \cite{Nientiedt1999, Stoyko2011}. These compounds have interlayer distances on the order of 10 $\mathrm{\AA}$, and magnetic interactions are thus expected to be highly-two-dimensional, thus theoretically enhancing quantum fluctuations. However, the rare-earth cations are octahedrally coordinated by pnictide anions, which act as strong crystal field field ligands and can have a substantial impact on the crystal field interactions of the 4\textit{f} electrons and result in strong anisotropies that can modify frustration. The impact of crystal field anisotropies on the magnetic properties of CeCd$_3$P$_3$ and CeCd$_3$As$_3$ have recently been explored, and these anisotropies have been suggested as a primary mechanism for the emergence of magnetic order in these systems below $T_N = 0.42$ K \cite{Uzoh2023}, despite the aforementioned two-dimensionality. 

    \begin{figure}[h]
        \includegraphics[width=0.48\textwidth]{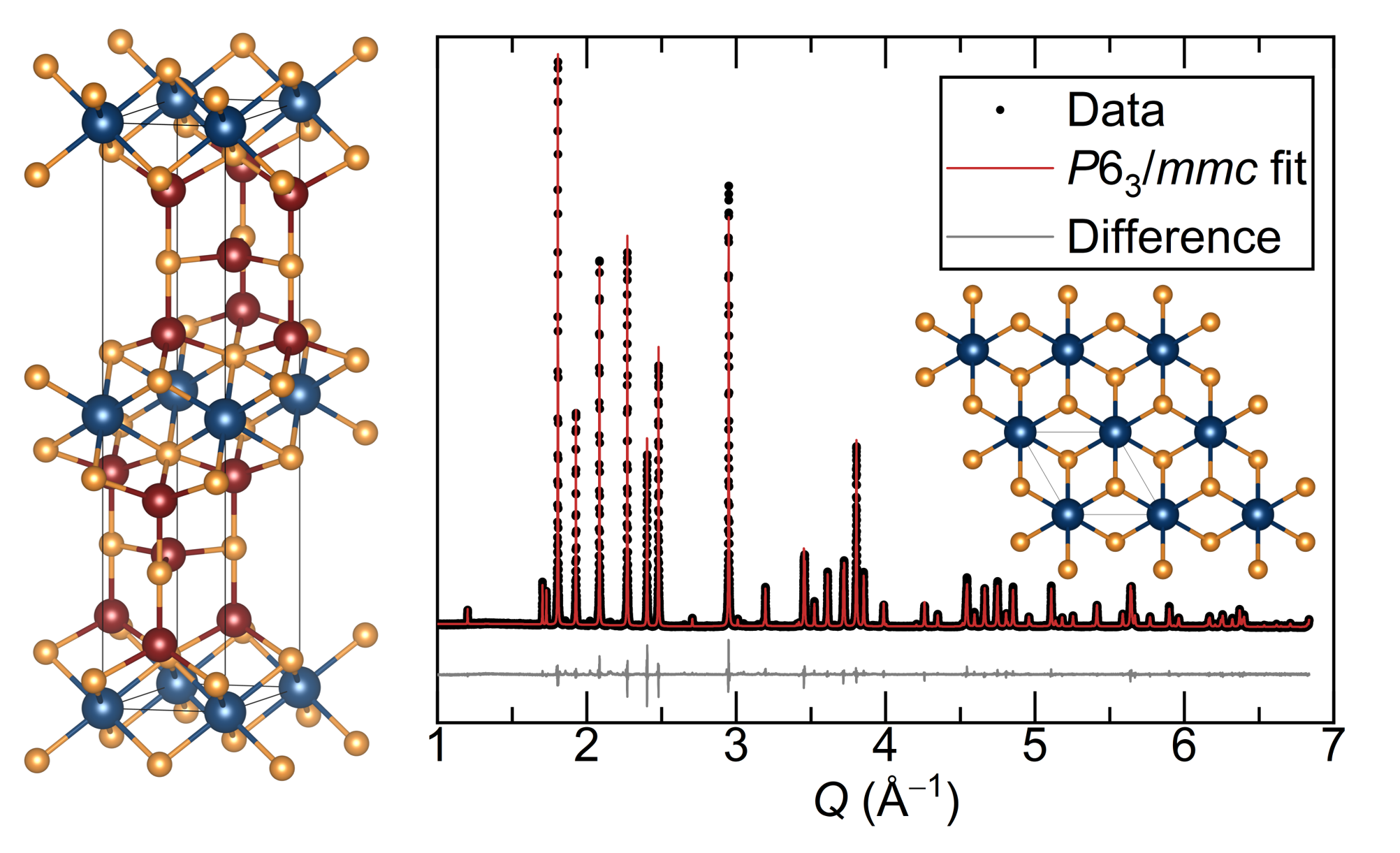}
        \caption{Crystal structure and synchrotron X-ray diffraction pattern of NdCd$_3$P$_3$, collected at \textit{T} = 300 K at 11-BM. Maroon atoms are Nd, blue atoms are Cd and orange atoms are P.}
    \end{figure}

The \textit{R}\textit{M}$_3$\textit{Pn}$_3$ family of materials is quite extensive, and many rare-earth variants (from \textit{R} = La to Er) have been found for the \textit{R}Zn$_3$P$_3$ subclass \cite{Nientiedt1999}. For the \textit{R}Cd$_3$P$_3$ subclass, only \textit{R} = La, Ce, and Pr have been reported \cite{Nientiedt1999,Lee2019}. Previous measurements on LaCd$_3$P$_3$ and CeCd$_3$P$_3$ demonstrated metallic states in single-crystalline samples with an unknown, possibly structural, transition at \textit{T} = 170 K and 126 K, respectively \cite{Lee2019}. In contrast, polycrystalline samples displayed insulating behavior \cite{Higuchi2016}. In addition, for the \textit{R}Zn$_3$P$_3$ compounds, an increasing tendency toward metallicity appears with increasing \textit{R} radius, which points toward a likely balance between unit cell volume and band dispersion near the Fermi level \cite{Kabeya2020}. Thus, exploring the \textit{R}Cd$_3$P$_3$ materials class as well as the overall \textit{R}\textit{M}$_3$\textit{Pn}$_3$ phase space may provide an avenue toward the systematic control of metallicity in these two-dimensional, naively frustrated, magnetic materials.

In this article, we report the structural and electronic properties of NdCd$_3$P$_3$, a new member of the \textit{R}Cd$_3$P$_3$ family of materials. We find that it is isostructural to previously-reported LaCd$_3$P$_3$ and CeCd$_3$P$_3$ variants.  We further report that NdCd$_3$P$_3$ is an electrical insulator with a charge transport gap of $E_G = 0.60$ eV, in agreement with density functional theory calculations predicting an insulating state. Despite the aforementioned large atomic distances between Nd$^{3+}$ triangular layers, our measurements of heat capacity and magnetic susceptibility reveal a magnetic ordering transition at \textit{T}$_C$ = 0.34 K, which is nearly equivalent to the mean-field magnetic interaction strength derived from fits to the susceptibility data to Curie-Weiss law, $\Theta_W$= - 0.39 K.

\section*{Methods}

Polycrystalline samples of \textit{R}Cd$_3$P$_3$ were prepared using conventional solid state synthetic techniques. First, lanthanide monophosphide \textit{R}P precursors were prepared by direct reaction of elemental rare earth powders (Fisher Scientific, 99.9\% REO) with red phosphorus (BeanTown Chemical, 99.9999\%). Stoichiometric amounts of the elements were weighed and ground together in an argon-filled glovebox, followed by heating in alumina crucibles at 850 $^{\circ}$C for 48 hours in sealed, evacuated silica ampoules. These precursors were then combined with a stoichiometric amount of Cd$_{3}$P$_{2}$ (Alfa Aesar, 99.5\% metals basis), and ball milled in a Spex 8000D Mixer/Mill for one hour using a tungsten carbide vial and tungsten carbide balls. The milled mixture was then sealed in evacuated silica tubes and heated to 800 $^{\circ}$C for 24 hours before being quenched in water.

High resolution synchrotron powder X-ray diffraction (XRD) data were collected at the 11-BM beamline at the Advanced Photon Source (APS) at Argonne National Laboratory, using an incident wavelength of 0.475799 $\mathrm{\AA}$. Data were collected at \textit{T} = 300 K. Powder XRD data were also collected using a laboratory x-ray source, a Panalytical Empyrean diffractometer employing Cu K$\alpha$ X-rays in a Bragg-Brentano geometry. Rietveld refinements of the data were performed using the TOPAS Academic software package. 

Magnetic susceptibility measurements were carried out in a Quantum Design Magnetic Property Measurement System (MPMS3), as well as using a vibrating sample mode (VSM) option on a Quantum Design 14 T Dynacool Physical Property Measurement System (PPMS). Resistivity and heat capacity measurements were performed using the PPMS using heat capacity and electrical transport options. Heat capacity measurements between \textit{T} = 0.1 and 2 K were performed using a Quantum Design dilution refrigerator.

First-principles electronic structure calculations based on density functional theory (DFT) were performed using the Elk code. This code is an all-electron, full-potential linearized augmented plane wave (LAPW) code with local orbitals (LO). Calculations were performed using the PBESol exchange functional \cite{Zhang1998}, and included spin-orbit coupling. Calculations for NdCd$_3$P$_3$ included a Hubbard $U$ term ranging from 4.5 - 6.5 eV, applied to the Nd$^{3+}$ $4f$ states. Calculations were also performed using the Vienna ab Initio Simulation Package (VASP) version 5.4.4. All calculations employed the Perdew-Burke-Ernzerhof (PBE) functional and projector-augemented wave potentials.

\section*{Results And Discussion}

\begin{figure}[h]
    \includegraphics[width=0.4\textwidth]{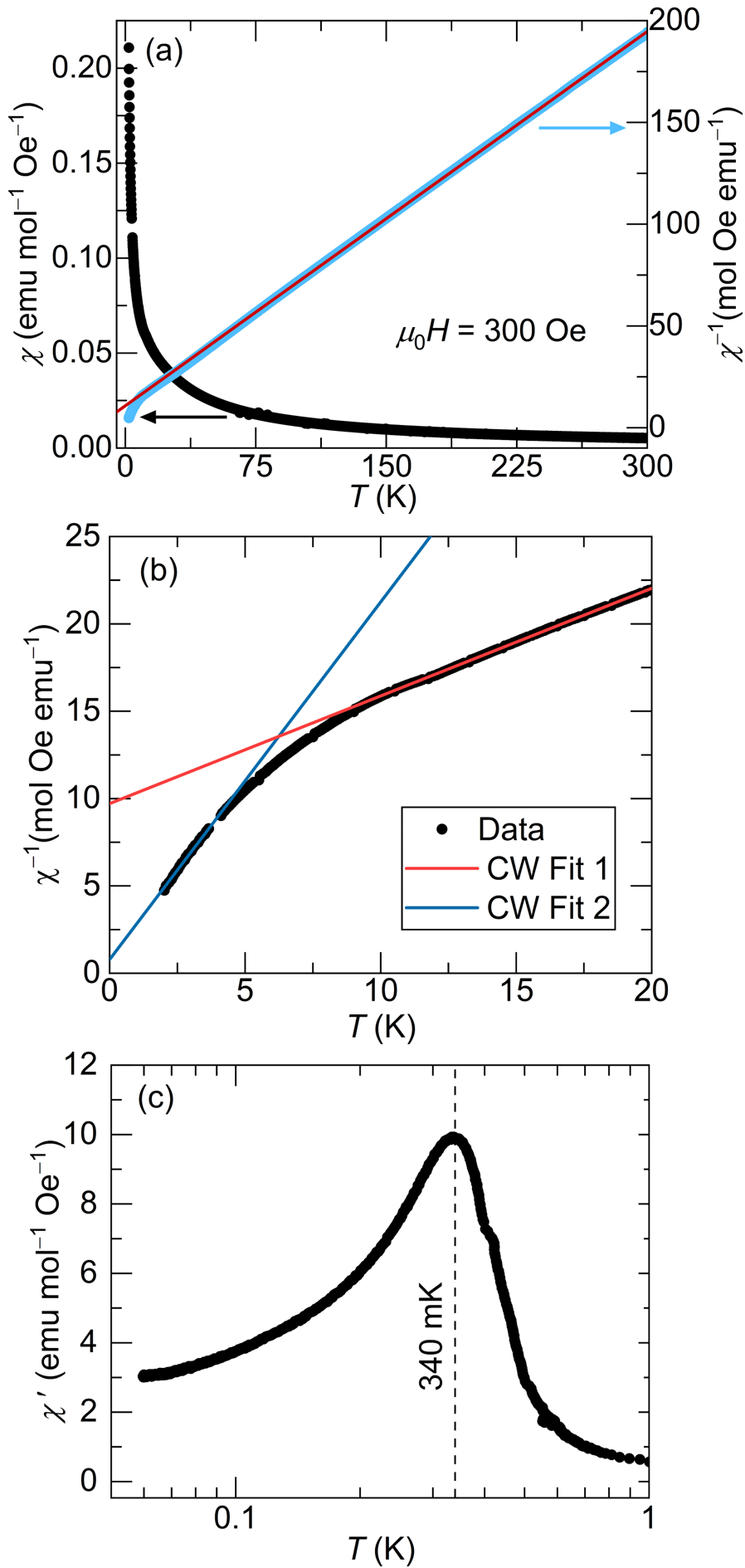}
    \caption{\textbf{A}. DC magnetic susceptibility of NdCd$_3$P$_3$, demonstrating full Curie-Weiss behavior between $T = 50 - 300$ K. Inset shows no bifurcation of zero-field cooled vs. field cooled data sets. \textbf{B}. Low temperature Curie-Weiss fits, demonstrating two regions, likely due to crystal field depopulation. \textbf{C}. AC susceptibility data collected between \textit{T} = 0.06 and 2 K, demonstrating an ordering transition centered around \textit{T} = 340 mK.}
\end{figure}

Synchrotron powder XRD data are shown in Figure 1, and structural parameters can be found in the supplementary information\cite{SuppInfo}. NdCd$_3$P$_3$ is isostructural to LaCd$_3$P$_3$ and CeCd$_3$P$_3$, which have been previously reported to crystallize in the hexagonal, $P6_3/mmc$ ScAl$_3$C$_3$ structure type \cite{Gesing1998}. NdCd$_3$P$_3$ is layered, and layers of edge-sharing NdP$_6$ octahedra are separated 10.46 $\mathrm{\AA}$ away from each other by Cd$_3$P$_3$ layers. These cadmium phosphide layers are built up of tetrahedral, corner-sharing CdP$_4$ units, and trigonal planar CdP$_3$ units. The large separation between magnetic Nd$^{3+}$ layers by many nonmagnetic atoms should naively promote a dominant two-dimensional exchange coupling within the triangular lattice planes.

\begin{figure}
    \includegraphics[width=0.48\textwidth]{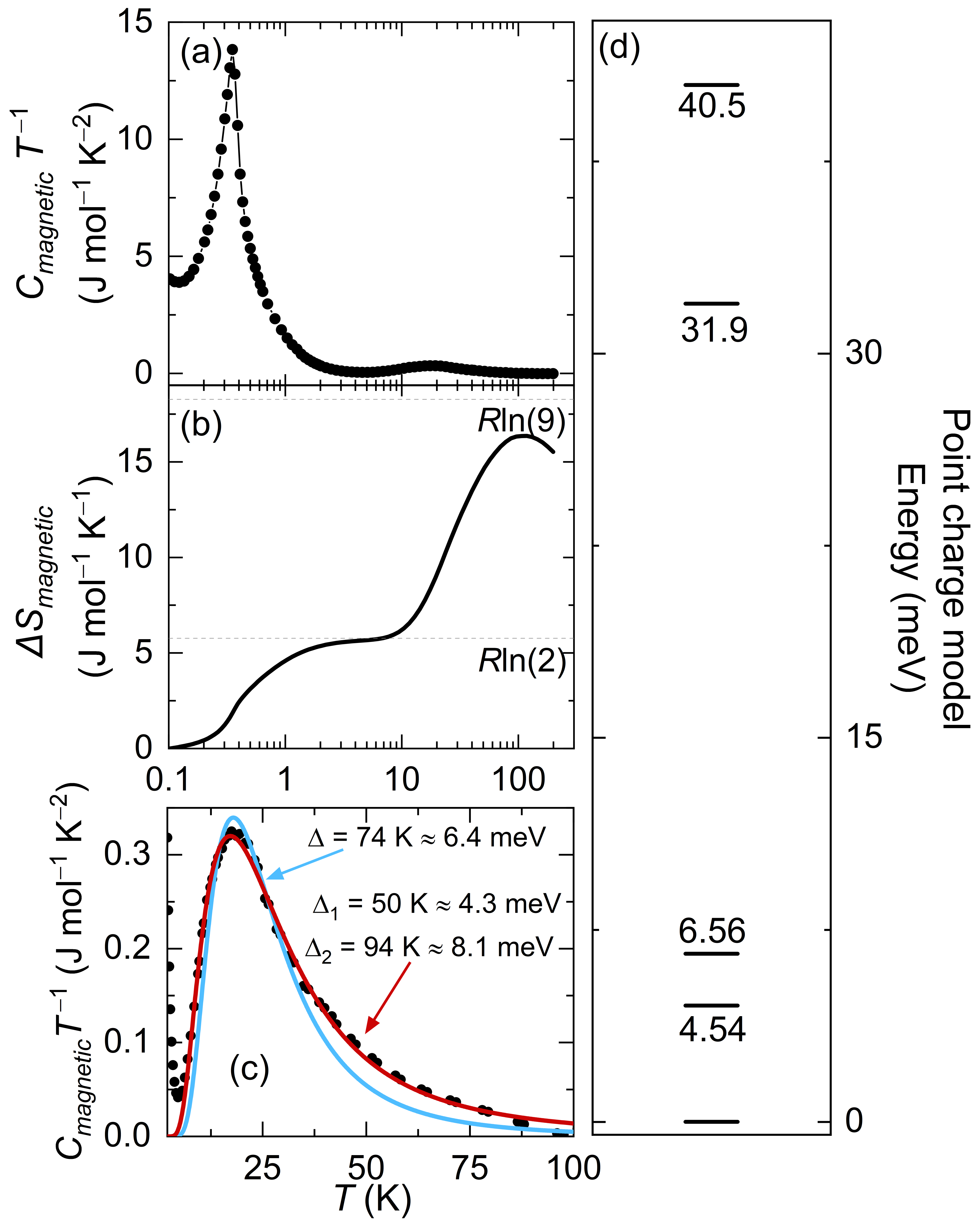}
    \caption{\textbf{A}. Magnetic heat capacity for NdCd$_3$P$_3$, extracted by subtracting a scaled LaCd$_3$P$_3$ diamagnetic analogue data set. \textbf{B}. Change in magnetic entropy of NdCd$_3$P$_3$, which plateaus near Rln(2) at low temperature before reducing toward zero at the ordering transition. \textbf{C}. Predicted crystal field levels as extracted using a point charge model.}
\end{figure}

The DC magnetic susceptibility of NdCd$_3$P$_3$ is shown in Figure 2. It follows Curie-Weiss behavior between \textit{T} = 50 - 300 K, and fits yield an effective, local Nd$^{3+}$ moment of $p_\mathrm{{eff}}$ = 3.62. The high temperature local moment is consistent with the expected $p_\mathrm{{eff}}$ = 3.62 for a free Nd$^{3+}$ $j = 9/2$ ion with a Land\'{e} $g = 8/11$. No irreversibility is observed between zero-field cooled (ZFC) and field-cooled (FC) data down to 2 K, implying an absence of any spin freezing in this temperature range. 

At low temperatures, the magnetic susceptibility deviates from the high-temperature, fixed moment Curie-Weiss behavior, likely due to the depopulation of low-lying Nd$^{3+}$ crystal field states, as shown in Figure 2 (b). The deviation starts below \textit{T} = 50 K, in agreement with heat capacity data discussed later in this paper. Once the low-lying excited states are depopulated upon cooling, the lowest temperature region can then be refit to a new Curie-Weiss model with an effective Nd$^{3+}$ moment of 1.98 $\mu_B$. In this low temperature regime, the mean field is antiferromagnetic and $\Theta_{W} = -0.38$ K, consistent with the ordering temperature observed in both AC susceptibility and heat capacity measurements as discussed below.

Figure 2 (c) displays the ac susceptibility data collected down to 60 mK.  Upon cooling below 2 K, the susceptibility continues to diverge in the condensed $S_{\mathrm{eff}}=1/2$ ground state and peaks at $T_N=340$ mK before rapidly decreasing.  This is consistent with an antiferromagnetic ordering transition arising at the same mean-field $\Theta_{W}$ value determined from the temperature Curie-Weiss fit in the crystal field ground state. No systematic frequency dependence can be observed in the ac susceptibility data, as demonstrated in the supplementary information \cite{SuppInfo}.

Heat capacity data were also collected on NdCd$_3$P$_3$ from \textit{T} = 80 mK to 300 K, and identical measurements were performed on LaCd$_3$P$_3$, which was then used to subtract off the lattice contribution.  Figure 3 (a) shows the remaining magnetic contribution to the heat capacity where two main features are observed: a peak with a maximum at \textit{T} = 340 mK, and a second, broader feature centered around \textit{T} = 18 K. The sharp peak represents a transition into a magnetically ordered state, as also suggested by the AC susceptibility data (see Figure 2 (c)). This ordering temperature is consistent with the extracted Weiss temperature of $\Theta_{W} = -0.38$ K. Based on the negative Weiss temperature, it suggests that this peak is a sign of an antiferromagnetic ordering transition.

Turning to the second, higher temperature peak in the magnetic heat capacity, the feature is highlighted in Figure 3 (c). This asymmetric peak can be fit with both a two-level and a three-level Schottky anomaly model, yielding gaps of $\Delta_{Schottky} = 6.4$ meV for the two-level and $\Delta_{Schottky}^1 = 4.3$ meV and $\Delta_{Schottky}^2 = 8.1$ meV for the three-level model. While both models have comparable statistics, a three-level Schottky anomaly both qualitatively fits the data better, and is in agreement with Nd$^{3+}$ point charge calculations shown in Figure 3 (c). These calculations were performed using the PyCrystalField package \cite{Scheie2021}, and assumed +3 charges for Nd cations and -3 for P anions, and employed the Rietveld-derived crystal structure of NdCd$_3$P$_3$ with experimental Nd-P bond distances. The integrated magnetic entropy is plotted in Figure 3 (b) and shows a gradual reduction of entropy on cooling across the Schottky anomaly and a plateau around $T = 10$ K at \textit{R}ln2, both consistent with an $S_{eff} = 1/2$ magnetic ground state.

\begin{figure}
    \includegraphics[width=0.4\textwidth]{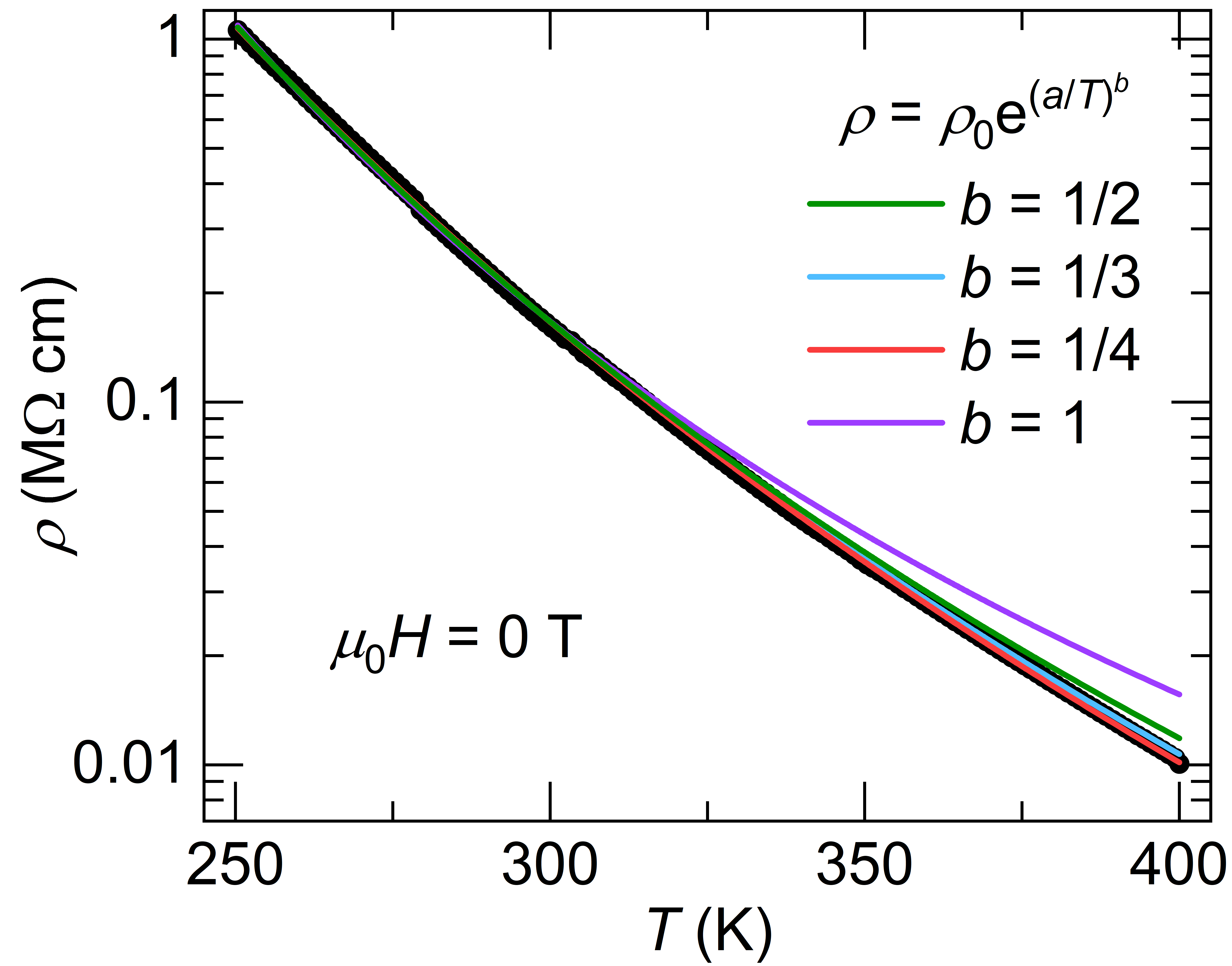}
    \caption{The measured electrical resistivity of NdCd$_3$P$_3$. The data was fit to various models of variable-range hopping transport, and found to be best fit using either $b = 1/3$ or $b = 1/4$ exponents, indicating either two- or three-dimensional hopping. The extracted band gap from the exponentially activated fit at lower temperatures (\textit{b} = 1) is \textit{E$_{gap}$} = 0.66(2) eV.}
\end{figure}

Figure 4 shows the electrical resistivity of a sintered pellet of NdCd$_3$P$_3$ in the range \textit{T} = 250 - 400 K. The resistivity increases with decreasing temperature, implying a bulk insulating state. The data was fit to a number of exponentially activated functions with varying \textit{b} exponents, which indicate either purely exponentially activated behavior ($b = 1$) or variable range hopping behavior ($b > 1$). In the case of NdCd$_3$P$_3$, the data is best fit to a variable range hopping model where the exponent can be fit to either $b = 1/3$ or $b = 1/4$, which represent two-dimensional and three-dimensional hopping, respectively \cite{Mott1969, Apsley1974}. The distinction between the two is made difficult due to the fact that the sample is polycrystalline. The NdCd$_3$P$_3$ transport gap \textit{E$_{gap}$} was extracted from a fit to the low temperature data using the exponentially activated Arrhenius form $\rho$= \,$\rho_0$e$^{a/2k_BT}$, where \textit{a} is the band gap and is on the order of 0.66(2) eV. This is consistent with trends from earlier work, which demonstrated polycrystalline samples of other \textit{R}Cd$_3$P$_3$ compounds to be insulators \cite{Higuchi2016}. In addition, \textit{E$_{gap}$} for NdCd$_3$P$_3$ is smaller than \textit{E$_{gap}$} for LaCd$_3$P$_3$ (0.73 eV) and CeCd$_3$P$_3$ (0.75 eV) \cite{Higuchi2016}. This follows the same trend as the \textit{R}Zn$_3$P$_3$ series of compounds, where the charge transport behavior goes from insulating to metallic across the lanthanides \cite{Kabeya2020}, and therefore the band gaps are expected to become smaller when employing higher atomic mass lanthanides. In contrast, however, single crystals of LaCd$_3$P$_3$ and CeCd$_3$P$_3$ show metallic behavior with a carrier density on the order of $10^{20}$ carriers/cm$^{3}$, as well as anomalies in the electrical transport behavior that are mirrored in the heat capacity \cite{Lee2019}. Neither the weakly metallic behavior nor the thermodynamic/transport anomalies reported in (La,Ce)Cd$_3$P$_3$ are observed in polycrystalline samples of NdCd$_3$P$_3$.

\begin{figure}
    \includegraphics[width=0.48\textwidth]{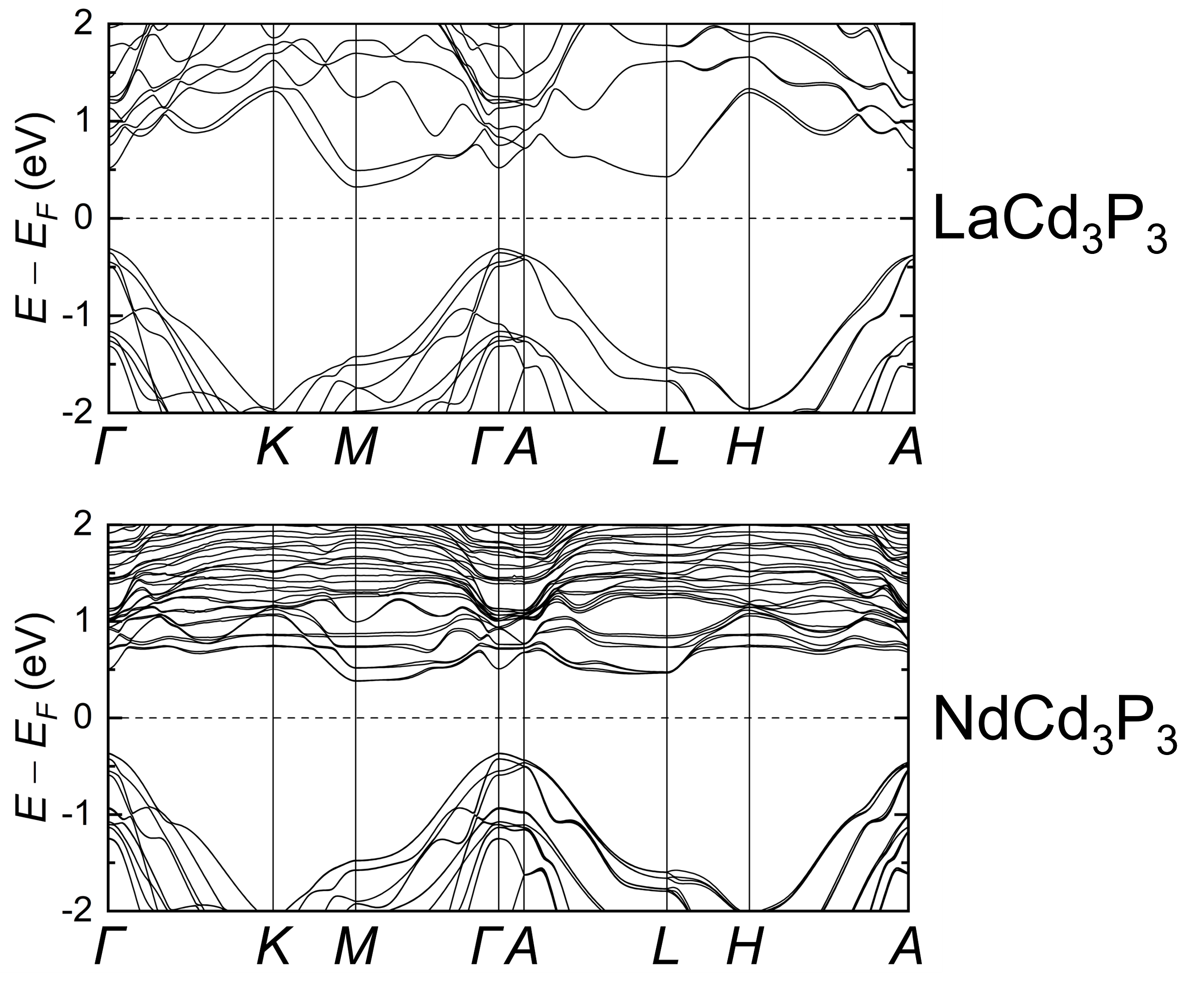}
    \caption{The electronic band structures of LaCd$_3$P$_3$ and NdCd$_3$P$_3$. Calculations include spin-orbit coupling. In both cases, the valence band maximum is made up nearly entirely of P-p states, whereas the conduction band minima are La or Nd $d$-bands. The Nd $f$-states can be found in the conduction band.}
\end{figure}

In order to understand the discrepancy in physical properties observed in NdCd$_3$P$_3$ compared to LaCd$_3$P$_3$ and CeCd$_3$P$_3$, we performed electronic band structure calculations using density function theory (DFT). The electronic band structures of LaCd$_3$P$_3$ and NdCd$_3$P$_3$ are shown in Figure 5. Spin-orbit coupling was included in these calculations, as well as a Hubbard U term for the Nd-\textit{f} states, U$_{\mathrm{eff}}$ = 5 eV. This is the lowest U$_{\mathrm{eff}}$ value required to fully gap out the Nd-4\textit{f} states that would otherwise be present at the Fermi level, as shown in the supplementary information for this manuscript \cite{SuppInfo}. 

Our calculations demonstrated that both LaCd$_3$P$_3$ and NdCd$_3$P$_3$ are gapped, with calculated indirect band gaps of 0.75 eV and 0.63 eV, respectively. These values are in rough agreement with the transport gaps derived from resistivity measurements \cite{Higuchi2016}, and suggest that the metallicity observed in studies of single crystals \cite{Lee2019} results from light doping. Given that the valence band maximum is primarily composed of disperse P $p$-bands and the conduction band minimum is composed of rather localized \textit{R} $d$-bands, the doped carriers in metallic samples are more likely to be holes. This is confirmed by a positive field-dependence of the Hall coefficient as previously reported in LaCd$_3$P$_3$ and CeCd$_3$P$_3$ \cite{Lee2019}.

\section*{Conclusion}

In this report, we synthesized polycrystalline samples of a new member of the \textit{R}Cd$_{3}$P$_{3}$ family of compounds, NdCd$_3$P$_3$. We find NdCd$_3$P$_3$ crystallizes in the ScAl$_{3}$C$_{3}$ structure and susceptibility data show that the compound realizes a $S_{\mathrm{eff}}=1/2$ magnetic ground state. Heat capacity and AC susceptibility measurements reveal the onset of magnetic order below \textit{T} = 340 mK at a temperature consistent with the antiferromagnetic Weiss field in this system. Charge transport data reveals that NdCd$_3$P$_3$ is insulating with an $E_{gap}$ of 0.66 eV, in good agreement with DFT calculations. 

\section*{Acknowledgments}

J.R.C. acknowledges support through the NSF MPS-Ascend Postdoctoral Fellowship (DMR-2137580). This research made use of the shared facilities of the NSF Materials Research Science and Engineering Center at UC Santa Barbara: NSF DMR-2308708. Use was made of computational facilities purchased with funds from the National Science Foundation (CNS-1725797) and administered by the Center for Scientific Computing (CSC), which is supported by the California NanoSystems Institute and the Materials Research Science and Engineering Center at UC Santa Barbara. This work was also supported by the National Science Foundation (NSF) through Enabling Quantum Leap: Convergent Accelerated Discovery Foundries for Quantum Materials Science, Engineering and Information (Q-AMASE-i): Quantum Foundry at UC Santa Barbara (DMR-1906325). This research used resources of the Advanced Photon Source, a U.S. Department of Energy (DOE) Office of Science user facility operated for the DOE Office of Science by Argonne National Laboratory under Contract No. DE-AC02-06CH11357.


\bibliography{biblio}

\end{document}